\documentclass[showpacs,preprintnumbers,amsmath,amssymb,preprint]{revtex4}

\usepackage{graphicx}
\usepackage{dcolumn}
\usepackage{bm}

\preprint{APS/superfocus}
\begin{document}
\title{Proposal for Subwavelength Optical Focusing}

\author{R. Gordon}
\affiliation{Department of Electrical and Computer Engineering, University of Victoria}
\email{rgordon@uvic.ca}

\begin{abstract}
Recently, radiationless electromagnetic interference (REI) has been considered to achieve focusing below the diffraction limit, with proposals for implementation in the microwave regime.
In this paper, subwavelength focusing using REI is
extended to the visible regime. It is shown that the highest-order mode of a
carefully designed metal-dielectric waveguide-array can provide the rapidly oscillating source field
for REI. Using finite-difference time-domain electromagnetic
simulations, an example structure is demonstrated. This structure
provides focusing to 0.21 of the optical wavelength at a distance of half the optical
wavelength from the waveguide-array. 
\end{abstract}

\pacs{78.66.Bz, 73.20.Mf, 42.25.Fx}

\maketitle
The ability to produce a sharp optical focus away from the focusing element is limited by the
decay of evanescent waves. The absence of these evanescent contributions leads to the fundamental diffraction limit.
Subwavelength focusing requires the amplification of evanescent waves that otherwise decay away
rapidly. One approach to achieving this in the optical regime is the ``poor-man's" perfect-lens~\cite{pendry2000}. In this scenario, the plasmon modes of a metal film effectively amplify the evanescent waves to produce an image at the
other side of the metal film. For each metal, the poor-man's perfect-lens approach works
around the specific wavelength where the surface plasmon dispersion curve flattens
out~\cite{pendry2000}. There have been several experimental demonstrations of this effect (for example, Refs.~\cite{melville2004,fang2005,melville2005}).

Recently, radiationless electromagnetic interference (REI) was proposed as another method to
create a subwavelength focus away from the focusing element~\cite{merlin2007}. This
proposal is geometric in nature and therefore not limited, at least in principle, to a specific wavelength. The basic idea of REI is to reconstruct the source field required at the focusing
element by propagating back from the desired focus spot distribution. This backward
propagation amplifies the evanescent (i.e., radiationless) modes associated with
fine optical features; therefore, the evanescent waves dominate focusing in the
subwavelength regime. The required source pattern oscillates rapidly in space~\cite{merlin2007}.

Since the initial proposal of radiationless electromagnetic interference, there have been several additional proposals for implementation and even experimental demonstrations within the microwave regime~\cite{wong2007,eleftheriades2008,grbic2008,markley2008}. It was
recognized that metal-dielectric structures would probably be required to extend these
results into the optical regime~\cite{merlin2007}; however, so far there have been no
proposals to achieve this goal.

In this work, I propose using a metal-insulator waveguide array to achieve the required
near-field source distribution for super-focusing in the optical regime. In particular,
the highest-order mode of a metal-dielectric waveguide array is considered because it
can provide the required rapid field oscillation. The waveguide is designed to support a magnetic
field pattern that best matches the desired source field of an ideal Gaussian focus.
To test this proposal, comprehensive finite-difference time-domain (FDTD) simulations are used to capture all the
effects of diffraction, including the excitation of spurious modes. These simulations show a 0.21 FWHM focus spot (only 80\% larger than the design specification) at half a wavelength from the termination of the waveguide array
structure.

Figure 1 shows, with a blue curve, the transverse magnetic field distribution at the focus for a Gaussian beam. The FWHM of the magnetic field intensity is 75~nm at the wavelength of $\lambda = 632.8$~nm. For this Gaussian focus, the plane-wave decomposition in the spatial-Fourier domain is also a Gaussian. Backward propagation to the source along $x$ is achieved by applying the operator $\exp(-i k_x x)$ to each plane-wave with transverse wave-vector $k_y$, where $k_x = \sqrt{4 \pi^2/\lambda^2 - k_y^2}$ is the wavevector along the $x$-direction. For $k_y > 2 \pi/\lambda$, this is an exponential increase in the plane-wave component; i.e., the evanescent waves are amplified. The red curve in Fig.~1 shows the near-field distribution after propagating backward by $\lambda/2$. This is the rapidly oscillating near-field that is required to produce the subwavelength focus. The magnetic field alternates sign between adjacent peaks.

\begin{figure}[htbp]
\centering\includegraphics[width=8cm]{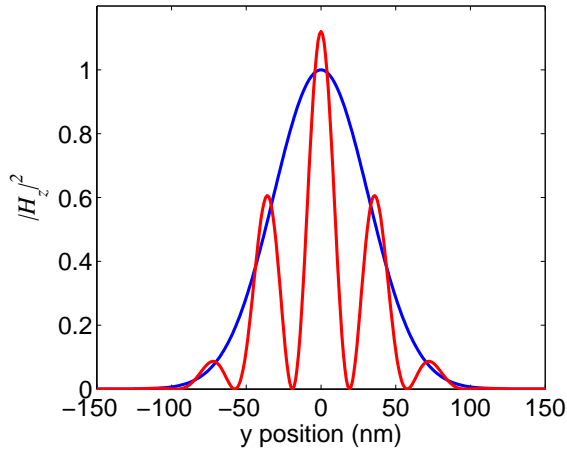}
\caption{(color online) Magnetic field intensity at the focus has a 74.5 nm FWHM Gaussian (blue curve). Magnetic field intensity at the source, $\lambda/2$ from the focus (red curve). $\lambda = 632.8$~nm. }
\end{figure}

Metal-dielectric-metal waveguides have a unique ability to confine light to transverse dimensions much smaller than the optical wavelength~\cite{economou1969}.
This is achieved because of the exponential decay of the electromagnetic field into the metal with negative relative permittivity. Here I consider an array of such metal-dielectric waveguides. The coupling in the array produces supermodes (e.g., Ref.~\cite{kapon1985}).

The highest-order supermode provides a very rapid (subwavelength) transverse oscillation. It is this type of rapid transverse oscillation that is required to produce an extreme subwavelength focus. It is not sufficient, however, to consider a uniform array, since this does not produce the desired subwavelength focus. Rather, the refractive index of each dielectric region should be altered slightly. (In an alternative proposal, the width of the guides can be varied as well to produce a higher effective index locally when the guide is made narrower -- this case is a straightforward extension of the results presented here.)

Figure 2 shows the highest order mode of a metal-dielectric waveguide array, designed to have a magnetic field distribution that is most similar to the red curve in Fig.~1. There are 9 dielectric layers. Each dielectric region is 10~nm wide and the refractive index values are 1.75, 1.73, 1.66, 1.56, 1.40, from the middle dielectric layer going outward along the $y$ direction. These refractive index values may be obtained with hybrid oxide layers. The center positions of the layers are 0~nm, $\pm$36~nm, $\pm$72~nm, $\pm$109~nm, and $\pm$147~nm, again going outward from the middle. The metal is gold which has a relative permittivity of $10.88 + i 0.80$ at 632.8~nm.


\begin{figure}[htbp]
\centering\includegraphics[width=8cm]{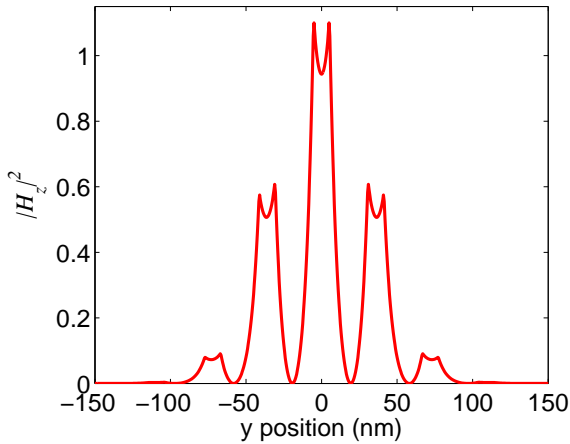}
\caption{(color online) Magnetic field intensity distribution of highest-order mode of metal-dielectric waveguide-array with 9 layers (outer layers have small electric field that is not visible). The refractive index values and the spacing of the dielectric regions are chosen to best-fit the source field distribution in Fig.~1.}
\end{figure}

Figure 3 shows the $x-$component of the Poynting vector magnitude, as calculated by comprehensive electromagnetic FDTD simulations. The source was the mode shown in Fig.~2. The metal-dielectric region is bounded by free-space for $x>0$. Convergence of the FDTD simulations was ensured by varying the simulation grid-size, and a 0.5~nm grid-size was used for the metal region. From the FDTD simulation, it is clear that there is the scattering at the interface to free-space. The scattering to very high spatial frequency components, as represented by the sharp-peak features of Fig.~2 (from, for example, the hyperbolic cosine-shaped points in the middle dielectric layer), decays away rapidly since they have a large transverse spatial frequency. The rapid decay ensures that these spurious evanescent modes do not effect the subwavelength focus at a half-wavelength distance from the interface. Due to the large mode-shape mismatch to the free-space propagating modes, the incident mode has near-perfect reflection. This ensures the strong excitation of radiationless evanescent waves without spurious propagating waves that would obstruct the subwavelength focus.

\begin{figure}[htbp]
\centering\includegraphics[width=12cm]{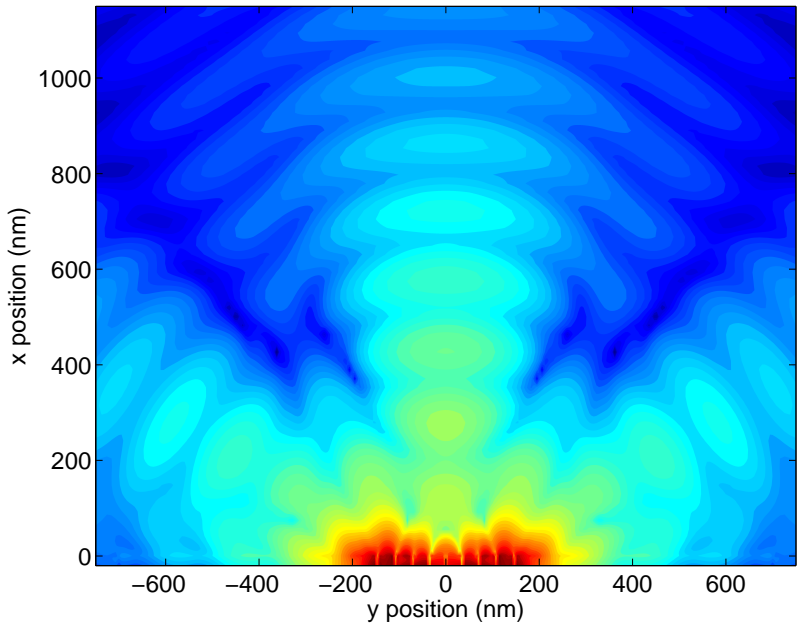}
\caption{(color online) Magnitude of Poynting vector, $x$-component, calculated by electromagnetic FDTD simulations. The colormap is plot on a logarithmic scale. The metal-dielectric waveguide array is present for $x<0$, and the region $x>0$ is free-space.}
\end{figure}

Figure 4 shows the transverse magnetic field intensity distribution at a half-wavelength distances from the metal-dielectric to air interface. A sharp subwavelength focus is observed, with a width that is only 80\% larger than the ideal case of Fig.~1. An important feature of this figure is that the side peaks are suppressed significantly below the main focus.

\begin{figure}[htbp]
\centering\includegraphics[width=8cm]{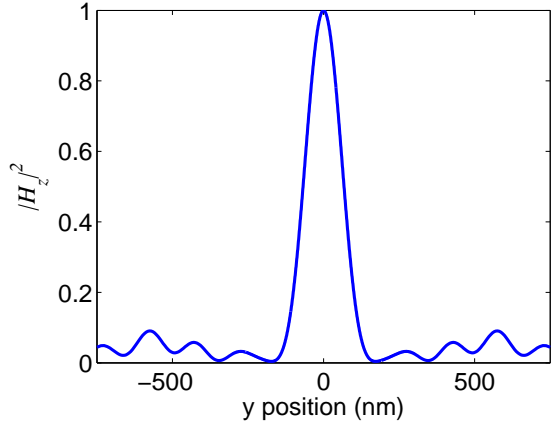}
\caption{(color online) Transverse magnetic field intensity at a half-wavelength away from free-space interface to the metal-dielectric waveguide array. The FWHM is 136~nm at the free-space wavelength of 632.8~nm. }
\end{figure}

In 2006, a 3~nm wide metal-dielectric-metal structure was made, and it was thinned to only 55~nm long~\cite{miyazaki2006}. The dimensions of that previous structure are smaller than the dimensions proposed here. Nevertheless, actual implementation will be a challenge since the proposed structure has several layers with varying dielectric constant. As mentioned previously, a single refractive-index dielectric can be used by varying the width of each of the dielectric layer. In that alternate approach, narrowing the layer modifies the effective index of the waveguide mode in that layer.

Physical demonstration will require the excitation of the highest-order waveguide mode. In one approach to excite that mode, several transverse conduits along the $y$-direction may be milled down, one to terminate at each channel. Such transverse conduits have been used to efficiently excite individual channel waveguides with propagation over several microns~\cite{dionne2006,lezec2007}. The spacing between the conduits can be chosen to impose the required phase-shift and the conduits can be separated by large distances in the $x$-direction to be easily addressed with conventional optics. Another approach would be to use a near-field probe at the focus of an array that is thinned down in the $x$-direction. Structures as thin as 55~nm have been demonstrated~\cite{miyazaki2006}. This probe would rely on reciprocity to excite the highest order mode. Further work is required in this area to determine the most feasible method of exciting the highest-order mode.

Finally, it is noted that the magnetic field decays substantially because of the evanescent nature of this technique. In particular, the magnetic field decays by two orders of magnitude in traveling from the interface at $x = 0$ to the focus at $x = 632.8$~nm. With high laser powers and high-dynamic range detectors that are commonplace today, this does not represent an insurmountable challenge. This rapid decay is also an exciting prospect to remote near-field mapping: it could provide an optical equivalent of electronic tunneling, since the field strength is exponentially distance dependent.

In conclusion, a proposal for extreme subwavelength focussing by REI was presented for optical wavelengths. In this proposal, the highest-order mode of a carefully designed metal-dielectric waveguide structure is used to produce the subwavelength focus with a considerable focal length. An example design showed a 136~nm FWHM magnetic field intensity focus over a half-wavelength focal length, for the wavelength of 632.8 nm. This proposal presents an exciting prospect to future applications in extreme subwavelength lithography, data storage and access, and imaging.


\begin{thebibliography}{14}
\expandafter\ifx\csname natexlab\endcsname\relax\def\natexlab#1{#1}\fi
\expandafter\ifx\csname bibnamefont\endcsname\relax
  \def\bibnamefont#1{#1}\fi
\expandafter\ifx\csname bibfnamefont\endcsname\relax
  \def\bibfnamefont#1{#1}\fi
\expandafter\ifx\csname citenamefont\endcsname\relax
  \def\citenamefont#1{#1}\fi
\expandafter\ifx\csname url\endcsname\relax
  \def\url#1{\texttt{#1}}\fi
\expandafter\ifx\csname urlprefix\endcsname\relax\def\urlprefix{URL }\fi
\providecommand{\bibinfo}[2]{#2}
\providecommand{\eprint}[2][]{\url{#2}}

\bibitem[{\citenamefont{Pendry}(2000)}]{pendry2000}
\bibinfo{author}{\bibfnamefont{J.~B.} \bibnamefont{Pendry}},
  \bibinfo{journal}{Physical Review Letters} \textbf{\bibinfo{volume}{85}},
  \bibinfo{pages}{3966} (\bibinfo{year}{2000}).

\bibitem[{\citenamefont{Melville et~al.}(2004)\citenamefont{Melville, Blaikie,
  and Wolf}}]{melville2004}
\bibinfo{author}{\bibfnamefont{D.~O.~S.} \bibnamefont{Melville}},
  \bibinfo{author}{\bibfnamefont{R.~J.} \bibnamefont{Blaikie}},
  \bibnamefont{and} \bibinfo{author}{\bibfnamefont{C.~R.} \bibnamefont{Wolf}},
  \bibinfo{journal}{Applied Physics Letters} \textbf{\bibinfo{volume}{84}},
  \bibinfo{pages}{4403} (\bibinfo{year}{2004}).

\bibitem[{\citenamefont{Fang et~al.}(2005)\citenamefont{Fang, Lee, Sun, and
  Zhang}}]{fang2005}
\bibinfo{author}{\bibfnamefont{N.}~\bibnamefont{Fang}},
  \bibinfo{author}{\bibfnamefont{H.}~\bibnamefont{Lee}},
  \bibinfo{author}{\bibfnamefont{C.}~\bibnamefont{Sun}}, \bibnamefont{and}
  \bibinfo{author}{\bibfnamefont{X.}~\bibnamefont{Zhang}},
  \bibinfo{journal}{Science} \textbf{\bibinfo{volume}{308}},
  \bibinfo{pages}{534} (\bibinfo{year}{2005}).

\bibitem[{\citenamefont{Melville and Blaikie}(2005)}]{melville2005}
\bibinfo{author}{\bibfnamefont{D.~O.~S.} \bibnamefont{Melville}}
  \bibnamefont{and} \bibinfo{author}{\bibfnamefont{R.~J.}
  \bibnamefont{Blaikie}}, \bibinfo{journal}{Optics Express}
  \textbf{\bibinfo{volume}{13}}, \bibinfo{pages}{2127} (\bibinfo{year}{2005}).

\bibitem[{\citenamefont{Merlin}(2007)}]{merlin2007}
\bibinfo{author}{\bibfnamefont{R.}~\bibnamefont{Merlin}},
  \bibinfo{journal}{Science} \textbf{\bibinfo{volume}{317}},
  \bibinfo{pages}{927} (\bibinfo{year}{2007}).

\bibitem[{\citenamefont{Wong et~al.}(2007)\citenamefont{Wong, Sarris, and
  Eleftheriades}}]{wong2007}
\bibinfo{author}{\bibfnamefont{A.~M.~H.} \bibnamefont{Wong}},
  \bibinfo{author}{\bibfnamefont{C.~D.} \bibnamefont{Sarris}},
  \bibnamefont{and} \bibinfo{author}{\bibfnamefont{G.~V.}
  \bibnamefont{Eleftheriades}}, \bibinfo{journal}{Electronics Letters}
  \textbf{\bibinfo{volume}{43}}, \bibinfo{pages}{1402} (\bibinfo{year}{2007}).

\bibitem[{\citenamefont{Eleftheriades and Wong}(2008)}]{eleftheriades2008}
\bibinfo{author}{\bibfnamefont{G.~V.} \bibnamefont{Eleftheriades}}
  \bibnamefont{and} \bibinfo{author}{\bibfnamefont{A.~M.~H.}
  \bibnamefont{Wong}}, \bibinfo{journal}{IEEE Microwave and Wireless Components
  Letters} \textbf{\bibinfo{volume}{18}}, \bibinfo{pages}{236}
  (\bibinfo{year}{2008}).

\bibitem[{\citenamefont{Grbic et~al.}(2008)\citenamefont{Grbic, Jiang, and
  Merlin}}]{grbic2008}
\bibinfo{author}{\bibfnamefont{A.}~\bibnamefont{Grbic}},
  \bibinfo{author}{\bibfnamefont{L.}~\bibnamefont{Jiang}}, \bibnamefont{and}
  \bibinfo{author}{\bibfnamefont{R.}~\bibnamefont{Merlin}},
  \bibinfo{journal}{Science} \textbf{\bibinfo{volume}{320}},
  \bibinfo{pages}{511} (\bibinfo{year}{2008}).

\bibitem[{\citenamefont{Markley et~al.}(2008)\citenamefont{Markley, Wong, Wang,
  and Eleftheriades}}]{markley2008}
\bibinfo{author}{\bibfnamefont{L.}~\bibnamefont{Markley}},
  \bibinfo{author}{\bibfnamefont{A.~M.~H.} \bibnamefont{Wong}},
  \bibinfo{author}{\bibfnamefont{Y.}~\bibnamefont{Wang}}, \bibnamefont{and}
  \bibinfo{author}{\bibfnamefont{G.~V.} \bibnamefont{Eleftheriades}},
  \bibinfo{journal}{Physical Review Letters} \textbf{\bibinfo{volume}{101}}
  (\bibinfo{year}{2008}).

\bibitem[{\citenamefont{Economou}(1969)}]{economou1969}
\bibinfo{author}{\bibfnamefont{E.~N.} \bibnamefont{Economou}},
  \bibinfo{journal}{Physical Review} \textbf{\bibinfo{volume}{182}},
  \bibinfo{pages}{539} (\bibinfo{year}{1969}).

\bibitem[{\citenamefont{Kapon et~al.}(1984)\citenamefont{Kapon, Katz, and
  Yarive}}]{kapon1985}
\bibinfo{author}{\bibfnamefont{E.}~\bibnamefont{Kapon}},
  \bibinfo{author}{\bibfnamefont{J.}~\bibnamefont{Katz}}, \bibnamefont{and}
  \bibinfo{author}{\bibfnamefont{A.}~\bibnamefont{Yarive}},
  \bibinfo{journal}{Optics Letters} \textbf{\bibinfo{volume}{9}},
  \bibinfo{pages}{125} (\bibinfo{year}{1984}).

\bibitem[{\citenamefont{Miyazaki and Kurokawa}(2006)}]{miyazaki2006}
\bibinfo{author}{\bibfnamefont{H.}~\bibnamefont{Miyazaki}} \bibnamefont{and}
  \bibinfo{author}{\bibfnamefont{Y.}~\bibnamefont{Kurokawa}},
  \bibinfo{journal}{Physical Review Letters} \textbf{\bibinfo{volume}{96}}
  (\bibinfo{year}{2006}).

\bibitem[{\citenamefont{Dionne et~al.}(2006)\citenamefont{Dionne, Lezec, and
  Atwater}}]{dionne2006}
\bibinfo{author}{\bibfnamefont{J.~A.} \bibnamefont{Dionne}},
  \bibinfo{author}{\bibfnamefont{H.~J.} \bibnamefont{Lezec}}, \bibnamefont{and}
  \bibinfo{author}{\bibfnamefont{H.~A.} \bibnamefont{Atwater}},
  \bibinfo{journal}{Nano Letters} \textbf{\bibinfo{volume}{6}},
  \bibinfo{pages}{1928} (\bibinfo{year}{2006}).

\bibitem[{\citenamefont{Lezec et~al.}(2007)\citenamefont{Lezec, Dionne, and
  Atwater}}]{lezec2007}
\bibinfo{author}{\bibfnamefont{H.~J.} \bibnamefont{Lezec}},
  \bibinfo{author}{\bibfnamefont{J.~A.} \bibnamefont{Dionne}},
  \bibnamefont{and} \bibinfo{author}{\bibfnamefont{H.~A.}
  \bibnamefont{Atwater}}, \bibinfo{journal}{Science}
  \textbf{\bibinfo{volume}{316}}, \bibinfo{pages}{430} (\bibinfo{year}{2007}).

\end{thebibliography}

\end{document}